\documentclass[aip,jcp,amsmath,amssymb,reprint]{revtex4-1}
\usepackage{graphicx}
\usepackage{amssymb}
\usepackage{amsmath}
\usepackage{color}
\usepackage[]{hyperref}
\usepackage[caption=false]{subfig}
\usepackage{bm}
\usepackage{upgreek}
\usepackage[multidot]{grffile}
\usepackage{float}
\usepackage{relsize}
\usepackage{rotating}
\usepackage{array}
\usepackage{tikz}
\usepackage{tabularx}
\usepackage[utf8]{inputenc}
\usepackage{changes}
\usetikzlibrary{backgrounds}
\usetikzlibrary{calc,decorations,decorations.text,decorations.pathmorphing,shapes.callouts,shapes.symbols}
\usetikzlibrary{decorations.pathreplacing}
\usepackage{changes}
\usepackage{standalone}

\tikzset{fontscale/.style = {font=\relsize{#1}}}

\begin{document}

\title{Interface Deformations Affect the Orientation Transition of Magnetic Ellipsoidal Particles Adsorbed at Fluid-Fluid Interfaces}

\author{Gary B. Davies}
\email{g.davies.11@ucl.ac.uk}
\affiliation{Centre for Computational Science, University College London, 20 Gordon Street, London WC1H 0AJ, United Kingdom.}

\author{Timm Kr\"uger}
\email{timm.krueger@ed.ac.uk}
\affiliation{Institute for Materials and Processes, Department of Engineering, University of Edinburgh, Mayfield Road, Edinburgh EH9 3JL, Scotland, United Kingdom.}
\affiliation{Centre for Computational Science, University College London, 20 Gordon Street, London WC1H 0AJ, United Kingdom.}

\author{Peter V. Coveney}
\email{p.v.coveney@ucl.ac.uk}
\affiliation{Centre for Computational Science, University College London, 20 Gordon Street, London WC1H 0AJ, United Kingdom.}

\author{Jens Harting}
\email{j.harting@tue.nl}
\affiliation{Department of Applied Physics, Eindhoven University of Technology, P.O. Box 513, 5600
MB Eindhoven, The Netherlands.}
\affiliation{Faculty of Science and Technology, Mesa+ Institute, University of
Twente, 7500 AE Enschede, The Netherlands.}

\author{Fernando Bresme}
\email{f.bresme@imperial.ac.uk}
\affiliation{Department of Chemistry, Imperial College London, London, SW7 2AZ, United Kingdom.}
\affiliation{Department of Chemistry, Norwegian University of Science and Technology, Trondheim, Norway.}

\begin{abstract}
Manufacturing new soft materials with specific optical, mechanical and
magnetic properties is a significant challenge. Assembling and manipulating colloidal particles
at fluid interfaces is a promising way to make such materials. We use lattice-Boltzmann simulations
to investigate the response of magnetic ellipsoidal particles adsorbed at liquid-liquid interfaces to
external magnetic fields. We provide further evidence for the first-order orientation phase transition predicted by Bresme and Faraudo [\textit{Journal of Physics: Condensed Matter} 19 (2007), 375110]. We show that capillary interface deformations around the ellipsoidal particle significantly affect the tilt-angle of the particle for a given dipole-field strength, altering the properties of the orientation transition. We propose scaling laws governing this transition, and suggest how to use these deformations to facilitate particle assembly at fluid-fluid interfaces. 
\end{abstract}

\maketitle

\section{Introduction}

Colloidal particles adsorb strongly at fluid-fluid interfaces. Detachment energies of spherical particles can be orders of magnitude greater than the thermal energy, $k_B T$.~\cite{binks_particles_2002,bresme_nanoparticles_2007}
This means that colloidal particles can attach irreversibly to interfaces, and hence stabilize emulsions better than surfactants, which are usually able to freely adsorb and desorb from an interface.~\cite{binks_particles_2002}
The shape and contact angle of the particle dictate how strongly it attaches for a given particle size: shapes that occupy smaller interface areas detach more easily than those occupying larger areas.~\cite{aveyard_particle_1996,faraudo_stability_2003,koretsky_1971,levine_stabilization_1989,tadros_vincent_1983}

Particles migrate to the interface to replace some fluid-fluid surface area with particle-fluid surface-area, reducing the free energy, $F_{\gamma} = \oint_{\partial A} \gamma \,\rm{d}A$, where $\gamma$ is the surface tension and $\partial A$ the interface area. Although surfactants adsorb to interfaces in a similar manner to particles and also lower the free energy by reducing the surface tension, surfactant stabilised emulsions and particle stabilised emulsions can behave very differently.

Once colloidal particles adsorb at a fluid-fluid interface, particle-particle interactions caused by competing hydrodynamic, electromagnetic and capillary forces can lead to particles self-assembling into materials with specific mechanical, optical, or magnetic properties.~\cite{Madivala2009b,ozin2009nanochemistry} Capillary interactions arise when particles deform the fluid interface. These interface deformations can be induced by external forces and torques such as gravity, or even by particle shape alone.~\cite{botto_capillary_2012}
Capillary interactions have attracted much interest in recent years for their
role in, for example, the self assembly of anisotropic particles at fluid
interfaces,~\cite{bresme_nanoparticles_2007,botto_capillary_2012,lehle_ellipsoidal_2008,loudet_how_2011,Guzowski2011c} the suppression of the coffee ring effect~\cite{yunker_suppression_2011} and the
Cheerios effect.~\cite{vella_cheerios_2005}

Advances in materials science have enabled the production of anisotropic particles with precise shapes, sizes, and electromagnetic properties.~\cite{snoeks_colloidal_2000}  Particles can also be manufactured with embedded ferromagnetic~\cite{zabow_ellipsoidal_2014} or (super)-paramagnetic dipoles~\cite{a,li_colloidal_2011} so that they are able to interact with external magnetic fields. This combination of particle shape and particle functionality, facilitated by the embedded dipoles, opens up a whole range of new ways to control particle self-assembly at fluid-fluid interfaces into two-dimensional structures.

~\citet{bresme_orientational_2007} investigated the behaviour of magnetic prolate spheroidal particles adsorbed at fluid-fluid interfaces under the influence of a homogeneous external magnetic field acting parallel to the interface normal, showing a very rich phenomenology.~\cite{faraudo_stability_2003,bresme_orientational_2007} In their analytical model, particles interact with the field via an embedded dipole moment directed along the particle long axis. Using classical capillarity theory and Monte-Carlo simulations, they found that the particle long axis aligns with the field but that the alignment is not continuous: The model predicts that for a critical dipole-field strength the particle \textit{flips} from a tilted orientation to a vertical one, with respect to the interface.  Monte-Carlo simulations showed good agreement with theory, but identifying the orientation transition was difficult given the presence of thermal fluctuations at small scales and the low activation free energy associated with the transition when the particles are small (nanometre range).~\cite{bresme_computer_2008,bresme_orientational_2007} 

Our article builds on this idea, and explores the physical behaviour arising from the interplay of external magnetic fields and capillary interactions induced by particle anisotropy. We employ a multi-component lattice-Boltzmann model~\cite{shan_lattice_1993,shan_simulation_1994}
with immersed rigid particles~\cite{ladd_numerical_1994,ladd_numerical_1994-1,ladd_lattice-boltzmann_2001,aidun98,aidun10,bib:jens-komnik-herrmann:2004,bib:jens-janoschek-toschi-2010b,jansen_bijels_2011,frijters_effects_2012,bib:jens-floriang-2013} to simulate prolate spheroidal
particles with embedded dipoles adsorbed at a liquid-liquid interface under the
influence of a magnetic field acting parallel to the interface normal. We confirm
that the predicted first order orientation transition exists and propose scaling laws governing this transition. We show that in the continuum limit, where colloidal particles are much larger than solvent particles, interface deformations around the particle play a significant role in the orientation of the
particle at the interface \\\indent
Finally, we show that these deformations are dipolar in nature and may lead to capillary interactions between particles when many particles are adsorbed at a fluid-fluid interface, with potential implications for the controlled assembly of particles at fluid-fluid interfaces. \\\indent
Our paper is organised as follows. Section \ref{sec:theory} discusses previous theoretical models describing magnetic particles adsorbed at fluid-fluid interfaces under the influence of an external magnetic field. Section \ref{sec:simmethod} details our simulation model and methods. We present the main results in section \ref{sec:results} and conclude the article in section \ref{sec:conclusions}.

\section{Theory}
\label{sec:theory}

\begin{figure}
	\begin{tikzpicture}[xscale=1,yscale=1]
		\draw [ultra thick,-latex] (1.25,0.0) -- (1.25,1.0);
		\node at (1.25,0.5) [label=above right:{$\mathbf H$}] {};

		\draw[rotate around={-55:(4.5,0.0)}] (4.5,0.0) ellipse (0.6cm and 1.8cm);
		\shade[ball color=green, rotate around={-55:(4.5,0.0)}] (4.5,0.0) ellipse (0.6cm and 1.8cm);

		\draw [dashed, rotate around={-55:(4.5,0.0)}, -] (4.5,-0.1) -- (4.5,-1.8);
		\draw [dashed, rotate around={-55:(4.5,0.0)}, -] (4.5,0) -- (3.9,0.0);
		\node at (3.9,0.7) [fontscale=2] {$R_{\perp}$};
		\node at (2.7,-1.15) [fontscale=2] {$R_{\|}$};
		
		\draw[dashed,blue,thick] (4.5,0) to (7.5,0);
		
		\draw[thick,-latex] (4.5,0) to (4.5,1.75);
		\node at (4.5,2.0) {$\mathbf n$};
		
		\draw[thick,red] ([shift=(0:0.5cm)]4.5,0) arc (0:90:0.5cm);
		\node at (4.9,0.65) [fontscale=2] {$\phi$};
		\node at (5.35,0.25) [fontscale=2] {$\psi$};
		
		\draw [decorate,decoration={brace,amplitude=6pt},rotate around={180:(4.5,0.0)},=180,xshift=-4pt,yshift=0pt]
(3.7,0.0) -- (5.6,0.0) node [black,midway,yshift=-0.4cm,fontscale=2] 
{ $A_{rm}$};
		
		\draw [thick, rotate around={-55:(4.5,0.0)}, -latex] (4.5,0) -- (4.5,2.2);
		\node at (6.55,1.45) {$\boldsymbol \mu$};
		
		\node at (7.0,0.75) [fontscale=1] {fluid 1};
		\node at (7.0,-0.75) [fontscale=1] {fluid 2};
		\draw[dashed,blue,thick] (1.5,0) to (7.5,0);
	\end{tikzpicture}
	
	\caption{(Color online) A particle with an embedded dipole, $\boldsymbol{\mu}$, under the influence of an external magnetic field, $\mathbf{H}$, directed parallel to the interface normal. The angle between the particle dipole axis and the magnetic field, $\phi$, is related to the angle between the dipole axis and the undeformed interface (dashed blue line), $\psi$, by $\psi = \pi/2 - \phi$, which we call the tilt angle. $R_{\perp}$ and $R_{\|}$ are the particle radii perpendicular and parallel to the particle symmetry axis, respectively. $\mathbf {n}$ is the interface normal vector. $A_{rm}$ is the area removed from the interface by the presence of the particle (Eq.~\eqref{eq:crossareas}).}
	\label{img:single_particle}
\end{figure}
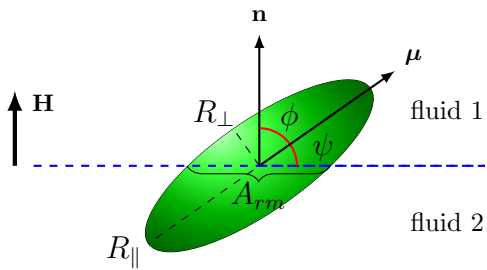

We consider the free energy of a prolate spheroidal particle with short and long axes $R_{\perp}$ and $R_{\|}$, respectively, adsorbed at a fluid-fluid interface under the influence of an applied external magnetic field, $\mathbf{H}$ (Fig.~\ref{img:single_particle}). In the absence of interface deformations around the particle and neglecting line tension effects, the free energy can be written as~\cite{bresme_orientational_2007,bresme_computer_2008,faraudo_stability_2003}
\begin{align}
\Delta F_{int} = -\gamma_{12} A_{rm} + (\gamma_{2p} - \gamma_{1p}) A_{2p} - B \cos \phi
\end{align}
\noindent where $B = \lvert \boldsymbol{\mu} \rvert \lvert \mathbf{H} \rvert$ represents the dipole-field strength, $\phi$ is the angle between $\boldsymbol{\mu}$ and $\mathbf{H}$ (Fig. \ref{img:single_particle}), $\gamma_{ij}$ is the surface-tension of the interface between phases 1 and 2 where $i,j =$ \{1: fluid $1$, 2: fluid $2$, p: particle\}. $A_{rm}$ is the area of the fluid-fluid interface removed by the presence of an adsorbed particle; $A_{1p}$ and $A_{2p}$ are the contact areas of the particle with fluid 1 and 2, respectively, and are related to the total particle surface area $A_p = A_{1p} + A_{2p} = 4 \pi R^2_{\perp} G(\alpha)$, where $G(\alpha) = \frac{1}{2} + \frac{1}{2} \frac{\alpha}{\sqrt{1-\alpha^{-2}}} \arcsin \sqrt{1-\alpha^{-2}}$ is a geometric aspect factor. 

The free energy $\Delta F_{int}$ above is calculated with respect to the free energy of the particle immersed in bulk phase $1$,  i.e., $\Delta F_{int} = F_{int} - F_{p1}$, where $F_{p1} = \gamma_{12} A_{12} + \gamma_{1p} A_p$, and $A_{12}$ is the total area of the unperturbed fluid-fluid interface.

For a neutrally wetting particle, $\gamma_{2p} = \gamma_{1p}$, and the free energy difference, henceforth referred to simply as the free energy, reduces to
\begin{align}
\label{eq:frenergy_simplified}
\Delta F_{int} = -\gamma_{12} A_{rm} - B \cos \phi,
\end{align}
where $A_{rm}$ is given by~\cite{bresme_orientational_2007,bresme_computer_2008}
\begin{align}
\label{eq:crossareas}
A_{rm} =  \frac{\pi R_{\|} R^2_{\perp}}{\sqrt{R_{\perp}^2 \cos^2\psi + R_{\|}^2 \sin^2 \psi}} 
\end{align}

The tilt angle, $\psi$, between the particle dipole-axis and the undisturbed interface is related to the dipole angle $\phi$ by $\psi = \pi/2 - \phi$ (Fig. \ref{img:single_particle}).

Substituting Eq.~\eqref{eq:crossareas} into Eq.~\eqref{eq:frenergy_simplified} gives the free energy, here made dimensionless by dividing by $A_p \gamma_{12}$, as a function of the tilt angle, $\psi$,  and the dipole-field strength, $B$:~\cite{bresme_orientational_2007}
\begin{align}
\label{eq:model}
\Delta \bar{F}_{int} &=  \frac{\Delta F_{int}}{A_p \gamma_{12}} = - \frac{B\cos(\pi /2 - \psi)}{A_p \gamma_{12}} \\\nonumber
& \quad - \frac{\alpha}{4G(\alpha)} \sqrt{\frac{1}{\cos^2(\psi) + \alpha^2 \sin^2(\psi)}}  \\\nonumber 
 \end{align}
\noindent where $\alpha = R_{\|}/R_{\perp}$ is the particle aspect ratio.

This thermodynamic model predicts that particles should experience a first-order orientation phase transition once a critical dipole-field strength, $B_c$, is reached.~\cite{bresme_orientational_2007, bresme_computer_2008} The model also predicts that the tilt angle, $\psi$, corresponding to a given dipole-field strength, $B$, depends on the aspect ratio, $\alpha$. Firstly, particles with larger aspect ratios require a larger dipole-field strength to transition from a tilted to a vertical state, with respect to the interface. Secondly, those particles transition at a smaller critical tilt angle, $\psi_c$, which is defined as the critical tilt angle corresponding to the critical dipole-field strength, $B_c$, at which the particle discontinuously transitions from a tilted to a vertical state. \\\indent
As the particle anisotropy tends to that of a spherical particle, $\alpha \to 1$, the first-order orientation phase transition disappears altogether. We reiterate that the model presented above in Eq.~\eqref{eq:model} assumes that the interface remains planar upon tilting, and therefore does not account for interface deformations around the particle. 

In the case of anisotropic particles such as ellipsoids, the particle curvature changes along the long particle axis and the three-phase contact-line must undulate around the particle in order for the contact-angle to remain constant, satisfying Young’s equation. Lattice Boltzmann (LB) simulations allow one to achieve a clear scale separation whereby the colloidal particle is much larger than the surrounding fluid solvent particles, approaching the continuum limit. Hence, LB simulations provide a promising approach to quantify the three-phase contact-line deformations as well as the impact of these deformations on the tilt angles and orientation transitions of the particles. We discuss our LB simulation approach in the following section.

\section{Simulation Model and Methods}
\label{sec:simmethod}

We employ the lattice-Boltzmann method on a
D3Q19 lattice\cite{bib:qian-dhumieres-lallemand} with the Shan-Chen
multi-component model\cite{shan_lattice_1993,aidun10} for the binary liquid part of the
system. Suspended particles are implemented following the pioneering work of Ladd and Aidun.\cite{ladd_numerical_1994,ladd_lattice-boltzmann_2001,aidun98,jansen_bijels_2011}
The LB method can be considered an alternative to traditional Navier-Stokes solvers for fluids and due to its local nature is well suited for implementation on supercomputers. 
We use the LB3D lattice Boltzmann code,~\cite{lb3d_citation} in which the above mentioned model is implemented. While elaborate descriptions of the model
implementation have been published
previously,\cite{jansen_bijels_2011,frijters_effects_2012,bib:jens-floriang-2013} we revise some relevant details for the present work in the following.

In the LB algorithm, each fluid component $c$ obeys the dynamical equation
 \begin{align}
\label{lbe}
f_i^c(\mathbf{x} + \mathbf{c}_i \Delta t, t + \Delta t) = f_i^c(\mathbf{x},t) + \Omega_i(\mathbf{x},t)
\end{align}
\noindent where  $i = 1,\ldots,19$ so that $f^c_i(\mathbf{x},t)$ represents the particle distribution function in direction $\mathbf{c}_i$ at lattice coordinate $\mathbf{x}$ and time $t$. $\Omega_i(\mathbf{x},t)$ is a generic collision operator: We use the Bhatnagar-Gross-Krook (BGK) operator~\cite{bhatnagar_model_1954,bib:benzi-succi-vergassola}
\begin{align}
\Omega_i = - {\frac{\Delta t}{\tau}} \left(f_i - f^\mathrm{eq}_i\right)
\end{align}
\noindent which has the effect of relaxing the system towards a local equilibrium distribution function $f^\mathrm{eq}_i$ on a time scale given by $\tau$.
The equilibrium distribution is a suitably chosen function of the fluid densities and velocities.
\noindent Apart from the common choice $\Delta x = \Delta t = 1$, \textit{i.e.} the introduction of ``lattice units'', we set $\tau = 1$ which leads to a numerical kinematic viscosity $\nu = \frac{1}{6}$ in lattice units.\\\indent
To simulate two immiscible fluids, we define two fluid components $c = 1, 2$ with densities $\rho^{(1)}$ and $\rho^{(2)}$. We also define an order parameter $\phi(\mathbf{x},t) = \rho^{(1)}(\mathbf{x},t) - \rho^{(2)}(\mathbf{x},t)$ which we call the ``colour'' of the fluid at a particular lattice site, $\mathbf{x}$.
We allow the fluids to interact via a mean-field force, the so called Shan-Chen approach.~\cite{shan_lattice_1993}\\\indent
To include immersed rigid particles, we use the method introduced by Ladd~\cite{ladd_numerical_1994,ladd_numerical_1994-1,ladd_lattice-boltzmann_2001} 
where particles are discretised on the lattice, and lattice sites occupied by a particle are treated as solid wall nodes in which bounce-back boundary conditions~\cite{chen_lattice_1998} are applied. For a more detailed description of how immersed rigid particles are handled in our simulations, we refer the reader to the relevant literature.~\cite{jansen_bijels_2011,frijters_effects_2012,bib:jens-floriang-2013,ladd_numerical_1994,ladd_lattice-boltzmann_2001,aidun98,aidun10} A noteworthy point is that our algorithm neglects thermal fluctuations. \\\indent
To investigate the response of magnetic ellipsoidal particles adsorbed at liquid-liquid interfaces to external magnetic fields, we first place a particle at a liquid-liquid interface with surface tension $\gamma_{12} = 0.06634$ in lattice units and allow the system to equilibrate. During equilibration, the interface diffuses to $\approx 5$ lattice sites wide. After equilibration, we apply a torque $\boldsymbol{\mathrm{T}} = \boldsymbol{\mu} \times \mathbf{H}$ to the particle, where $\mathbf{H}$ is a magnetic field acting parallel to the interface normal, $\mathbf{n}$ (Fig.~\ref{img:single_particle}).

\section{Results}
\label{sec:results}
\begin{figure}
\includegraphics[width=\linewidth]{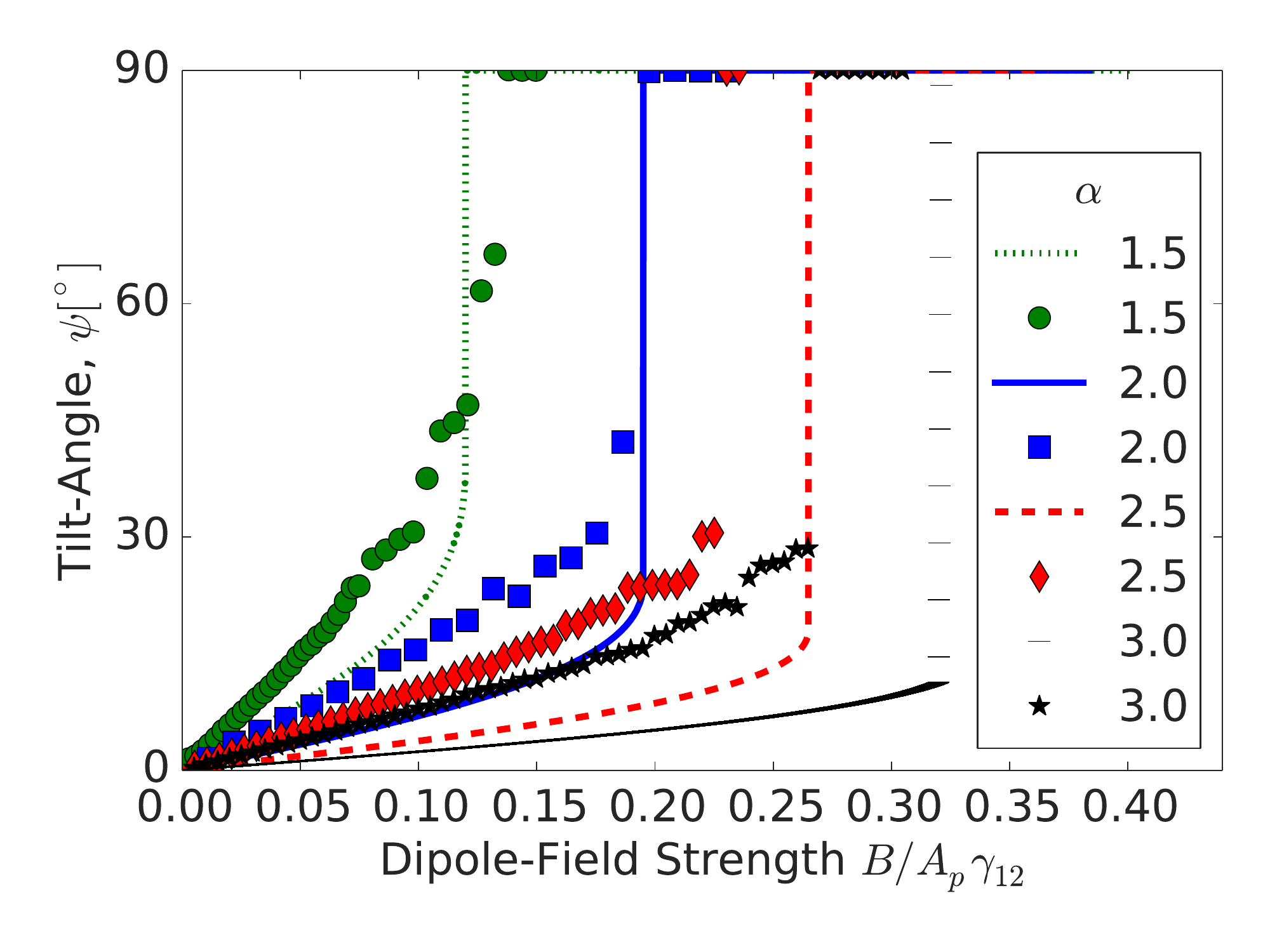} \\ [-0.4cm]
\caption{(Colour online) Comparison of $\psi(B)$ as obtained from the thermodynamic model in Eq.~\eqref{eq:model} (lines) and our numerical simulations (symbols) for various aspect ratios, $\alpha$. The dipole-field strength is made dimensionless by dividing by $A_p \gamma_{12}$. Although the numerical data qualitatively confirm the model predictions, in particular that the discontinuous transition exists, there are quantitative deviations.}
\label{plot:theory_vs_data}
\end{figure}

In Fig.~\ref{plot:theory_vs_data} we compare our simulation results with the predictions of the theoretical model in Eq.~\eqref{eq:model}.~\cite{bresme_orientational_2007} The LB simulations reproduce the theoretical predictions, namely,  we find a discontinuous transition of the tilt angle when the dipole-field strength is increased past a critical value, $B_c$. We also find that increasing $\alpha$ leads to a decrease of the tilt angle for a given dipole-field strength. Additionally, the critical dipole field strength and critical tilt-angle increases and decreases, respectively, as the aspect ratio is increased.

The LB simulations feature quantitative differences with the theoretical model in Eq.~\eqref{eq:model} that increase with the particle aspect ratio. The LB tilt angle is larger than the thermodynamic prediction, while the critical dipole-field strength becomes progressively smaller than the theoretical result as the aspect ratio increases (Fig.~\ref{plot:theory_vs_data}). These deviations are in contrast with the results obtained using atomistic Monte Carlo simulations of ellipsoidal particles adsorbed at fluid-fluid interfaces.~\cite{bresme_orientational_2007,bresme_computer_2008} In those simulations, the agreement between theory and simulation was quantitative for small aspect ratios, $\alpha = 1.2$, and low dipole-field strengths.~\cite{bresme_orientational_2007, bresme_computer_2008}

Interestingly, for larger particle aspect ratios, $\alpha=1.5$ and $2$ the atomistic simulations~\cite{bresme_orientational_2007, bresme_computer_2008} featured systematic deviations from the theory, similar to those observed in our LB simulations, although the simulated angles are much closer to the theoretical predictions than in our LB simulations. The particles investigated in the atomistic simulations were small, typically 5-10 times larger than the solvent particle diameter, approximately 1-3~nm. At this nanometre scale the granularity of the solvent is relevant, in contrast with our LB simulations, where the solvent particles approach the continuum limit. 
 \begin{figure}
\includegraphics[width=\linewidth]{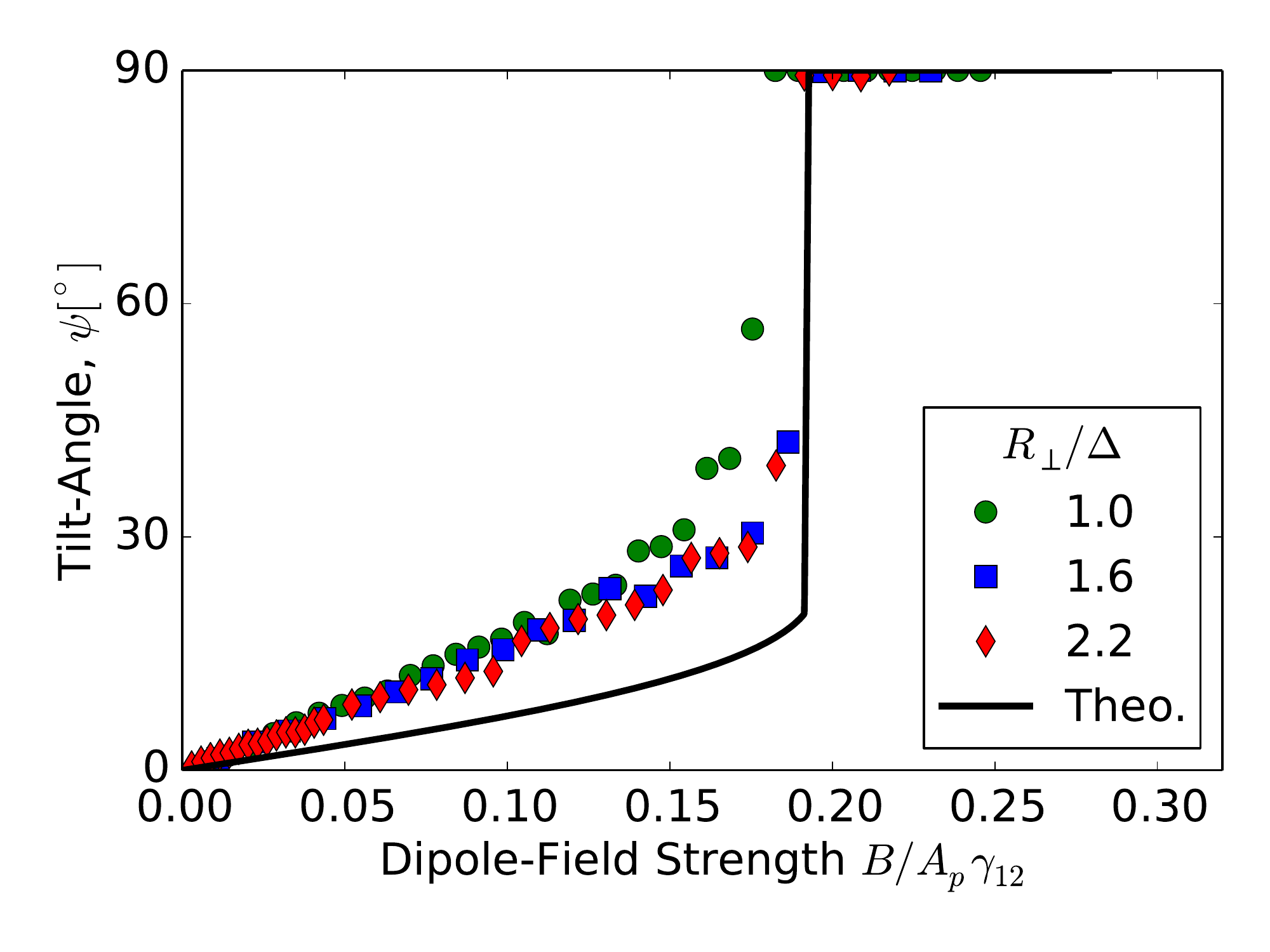} \\ [-0.4cm]
\caption{(Colour online) Convergence study. Numerical data (symbols) are compared with theoretical predictions (solid black line) for $\alpha = 2$. The ratio of the particle radius orthogonal to the symmetry axis, $R_{\perp}$ (Fig. \ref{img:single_particle}), to the interface width, $\Delta$, is varied. A ratio of $1.6$ is enough to ensure numerical convergence.}
\label{plot:interface_thickness}
\end{figure}

We first rule out the effect of the LB discretization as a possible explanation for the observed deviations between LB simulation and atomistic simulations by performing a grid refinement study, shown in Fig.~\ref{plot:interface_thickness}. By increasing the ratio of particle radius to interface width, $R_{\perp} / \Delta$ (where $\Delta \approx 5$ is the constant interface width~\cite{frijters_effects_2012}) for a particle with aspect ratio $\alpha = 2$, we show that the ratio $R_{\perp} / \Delta = 1.6$ is sufficient as the result for $R_\perp / \Delta = 2.2$ does not improve the agreement. This is consistent with previously published results.~\cite{jansen_bijels_2011} We ensured that the ratio $R_{\perp}/\Delta$ was at least $1.6$ in all our simulations.

\begin{figure}[t]
	
	 \subfloat[Simulation visualisation\label{img:single_p_contactline_render}]{
	\includegraphics[width=\linewidth]{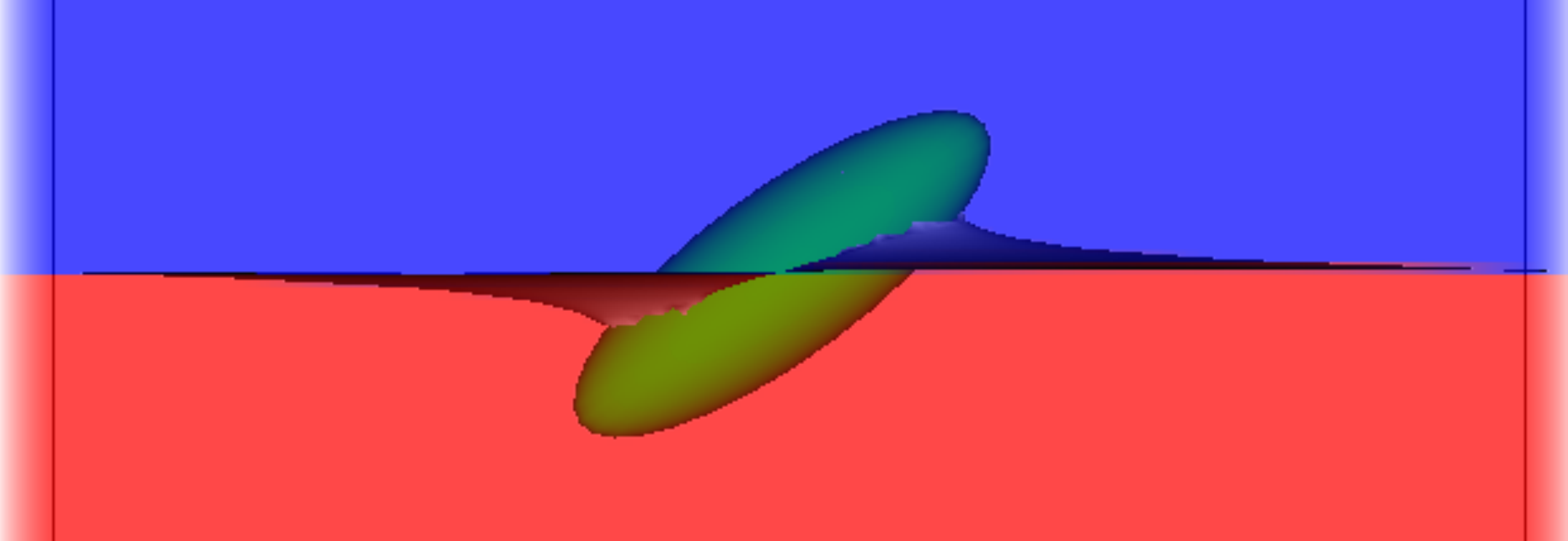} 
	}\\ 

\subfloat[Schematic illustration\label{img:single_p_contactline}]{
	\includegraphics[width=\linewidth]{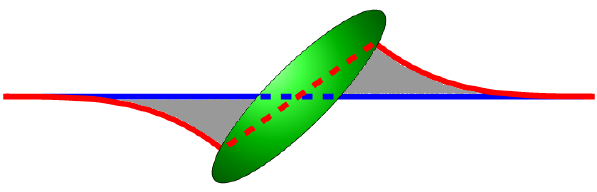}
	} \\ 

	\caption{(Colour online) (a) Simulation with aspect ratio $\alpha = 3$ under the influence of a dipole-field $B/A_p \gamma_{12} = 0.25$. We see a repositioning of the three-phase contact line compared with a flat interface. (b) Schematic showing the difference between the theoretically predicted area removed, $A_{rm}$ (blue dashed), and interface profile (blue solid), compared with the observed area removed, $A_{rm,d}$ (red dashed) and interface profile (red solid).}
	
	\label{imgs:interface_deforms}
\end{figure}

In the following, we consider the physical origin of the higher tilt angles predicted by LB as compared with the thermodynamic theory and Monte Carlo simulations.~\cite{bresme_orientational_2007,bresme_computer_2008} For ellipsoidal particles that do not feature a constant curvature along the three-phase contact-line, Young's equation dictates that the three-phase contact line should undulate around the particle. The surrounding fluid interface will be deformed as a consequence. We find that as a response to the external field and particle rotation, the interface deforms in an \textit{anti-symmetric}, dipolar fashion: The particle depresses the interface on one side and raises it on the other, as presented in Fig.~\ref{img:single_p_contactline_render} and sketched in Fig.~\ref{img:single_p_contactline}. Since the particles are neutrally wetting, quadrupolar interface deformations are absent.

The contribution of the interface deformation to the free energy may be included by rewriting Eq.~\eqref{eq:frenergy_simplified} as

\begin{align}
\label{newfreeen}
\Delta F_{int,d}(\psi) = -\gamma_{12} \left(A_{12} -  A_{12,d}(\psi)\right) - B \cos\left( \frac{\pi}{2} - \psi\right) 
\end{align}
\noindent
where $A_{12,d}(\psi)$ is the total deformed fluid-fluid interface area when a particle is adsorbed at the interface. $A_{rm,d} = A_{12} -  A_{12,d}$ is the area removed by the particle when it absorbs at the interface, including the effect of interface deformations. $A_{rm,d}$ converges to $A_{rm}$ (Eq.~\eqref{eq:crossareas}) in the flat interface limit, i.e. $\psi=0^{\circ}$ or $90^{\circ}$, and Eq.~\eqref{eq:model} is therefore recovered. The equilibrium tilt angle for a given dipole-field strength can then be obtained by minimizing $\Delta F_{int,d}$ with respect to $\psi$. 

\begin{figure}
	\includegraphics[width=\linewidth]{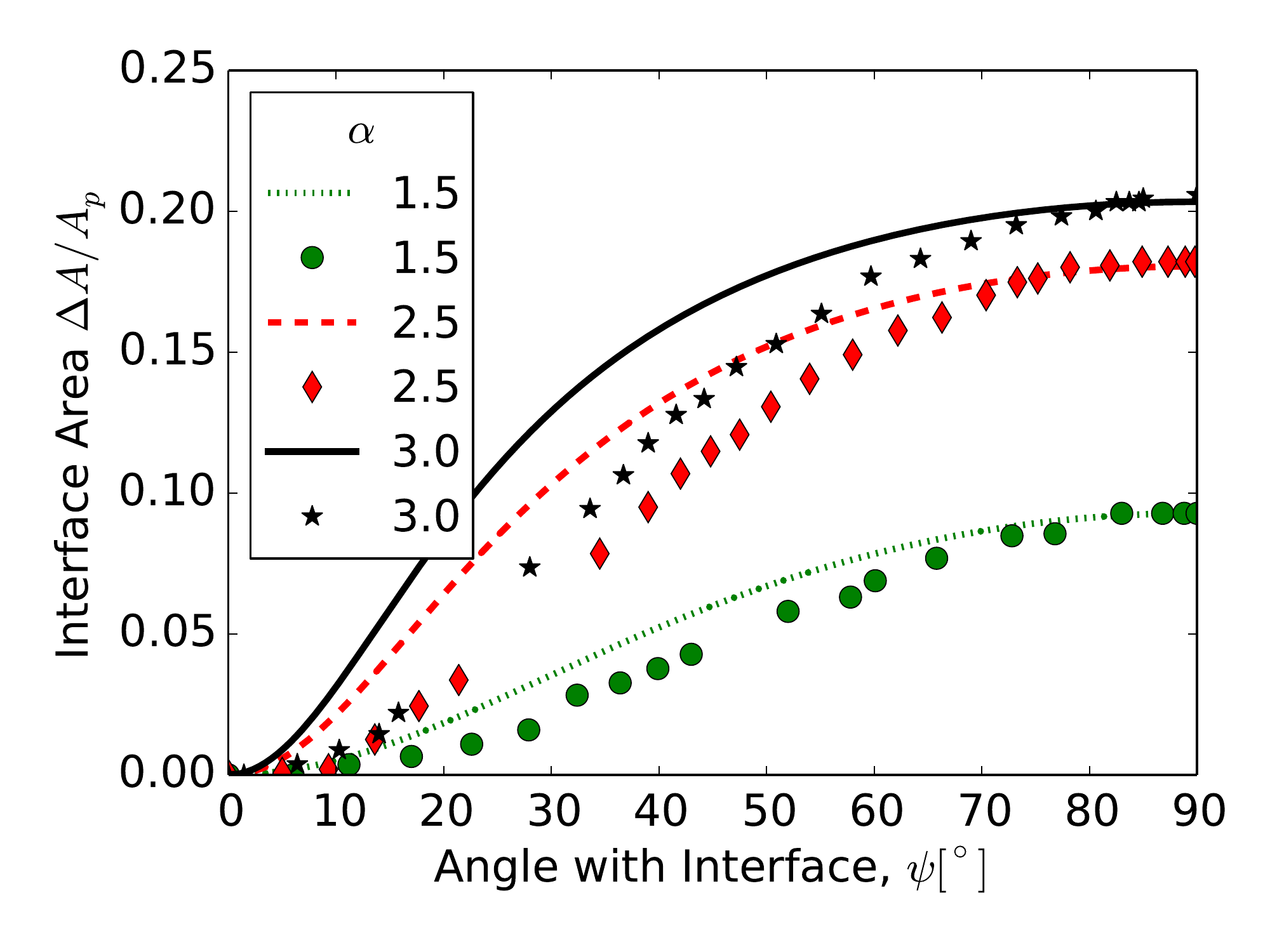} \\ 
	\caption{(Colour online) The difference between the predicted lines, $\Delta A =  A_{rm}(\psi=0) -  A_{rm}(\psi)$ and simulated symbols, $\Delta A =  A_{rm,d}(\psi=0) -  A_{rm,d}(\psi)$, is denoted by the grey shaded region in Fig.~\ref{img:single_p_contactline}. For $\psi = 0^{\circ}$ and $90^{\circ}$, there is no particle induced interface deformation and so the symbols and lines match.}
	\label{plot:interface_deform_plot}
\end{figure}
\begin{figure}
	\includegraphics[width=\linewidth]{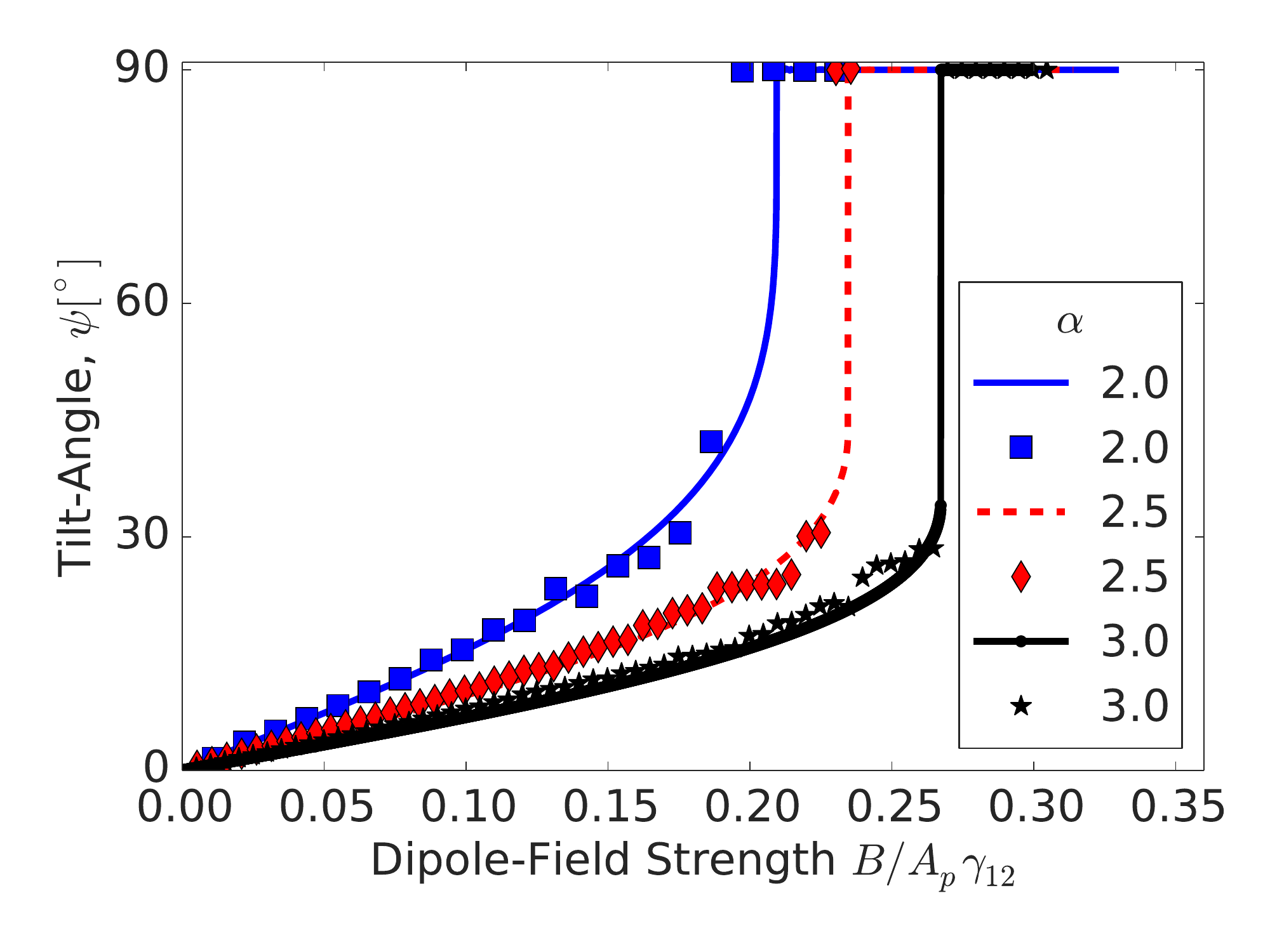} \\ 
	\caption{(Colour online) Prediction of the particle tilt-angle, $\psi$, including the effect of interface deformations. The lines represent the equilibrium angle corresponding to the minimum in the free energy presented in Eq.~\eqref{newfreeen}. The good agreement between the model (lines) and the data (symbols) suggests that interface deformations are the sole cause of the deviations observed in Fig.~\ref{plot:theory_vs_data}.}
	\label{plot:fitted}
\end{figure}
Figs.~\ref{img:single_p_contactline_render} and~\ref{img:single_p_contactline} show that the interface deformation removes more interface area than the planar approximation (see dashed lines in Fig.~\ref{img:single_p_contactline}), which lowers the free energy. At the same time, more interface area is generated as a result of the curvature of the deformed interface around the particle (see solid lines in Fig.~\ref{img:single_p_contactline}), which increases the free energy. It is the balance between these effects that determines the observed interface profile. However, the net effect is a decrease in total interface area.
In order to assess the impact of the removed interface area observed in our simulations, $A_{rm,d} = A_{12} - A_{12,d}$, we computed $A_{12,d}$ by fixing the particle orientation (and hence its tilt-angle) and integrating the area of the deformed interface when the particle is adsorbed at the interface. $A_{12,d}$ includes the contributions of both the increased interface area due to curvature around the particle and decreased interface area due to the contact-line re-arrangement, as described above. Therefore, $A_{rm,d}$ is the area removed by the particle including the additional curvature-induced interface area.

In Fig.~\ref{plot:interface_deform_plot}, we plot $\Delta A(\psi) = A_{rm,d}(\psi\!=\!0) - A_{rm,d}(\psi)$ obtained from our simulations (symbols) and compare with $\Delta A(\psi) = A_{rm}(\psi\!=\!0) - A_{rm}(\psi)$ calculated from the theoretical model in Eq.~\eqref{eq:crossareas} (lines). The difference between these two curves is denoted by the grey shaded region in Fig.~\ref{img:single_p_contactline}.

We see that the degree of interface deformation increases with particle aspect ratio. At small particle aspect ratios, $\alpha = 1.5$, where the particle is nearly spherical, the increased area due to the deformation of the interface is small. In this case, we expect that the deviations between the LB results and the analytical model are smaller. This conclusion is consistent with the dependence of the tilt angles reported in Fig.~\ref{plot:theory_vs_data}.

We also find that the deformed interface area with the particle adsorbed, $A_{12,d}$, is smaller than predicted by the theoretical model in Eq.~\eqref{eq:crossareas} for a given dipole-field strength, $B$. This smaller total interface area implies a lower free energy and hence a higher tilt-angle for a given dipole-field strength, which is also in agreement with the tilt angles reported in Fig.~\ref{plot:theory_vs_data}. For $\psi = 0^{\circ}$ and $90^{\circ}$, the curves converge, since for these tilt-angles there is no interface deformation for neutrally wetting particles. 

The interface area removed by the particle that we observed in our simulations, $A_{rm,d}$, can be used to recalculate the equilibrium tilt angle. We note however that there is not a simple analytical expression for $A_{rm,d}$, hence, we fitted our numerical areas to a function $A_{rm,d} = c + \frac{\pi a b^2}{\sqrt{a^2 \cos^2\psi + b^2 \sin^2 \psi}}$, where $a$, $b$, and $c$ are free parameters. We chose this functional form, taking inspiration from Eq.~\eqref{eq:crossareas}, since it captures the phenomenology of the observed discontinuous orientation transition. Using this fitted area, we then minimized Eq.~\eqref{newfreeen} with respect to the tilt-angle $\psi$ to obtain a new estimate of the tilt angles as a function of the dipole-field strength, $B$, shown in Fig~\ref{plot:fitted}.
\begin{figure}
\includegraphics[width=\linewidth]{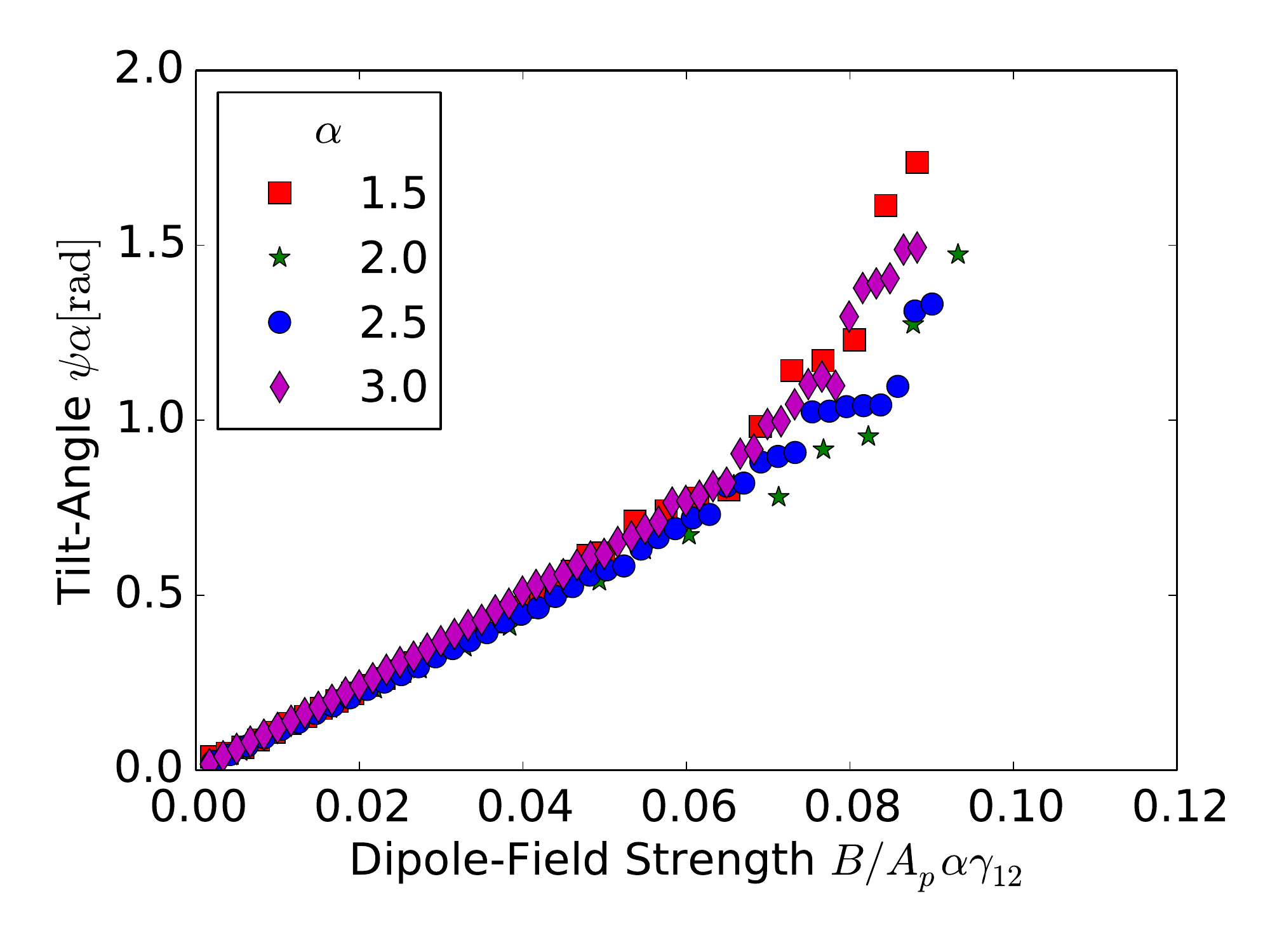} \\ [-0.4cm]
\caption{(Colour online) Tilt angle, $\psi\alpha$, \textit{vs.}~reduced dipole-field strength, $B / A_p \gamma_{12} \alpha$. Using this scaling, we observe a full collapse of the simulated data to a single master curve.}
\label{plot:alpha_scaling_law}
\end{figure}

\begin{figure}
	\includegraphics[width=\linewidth]{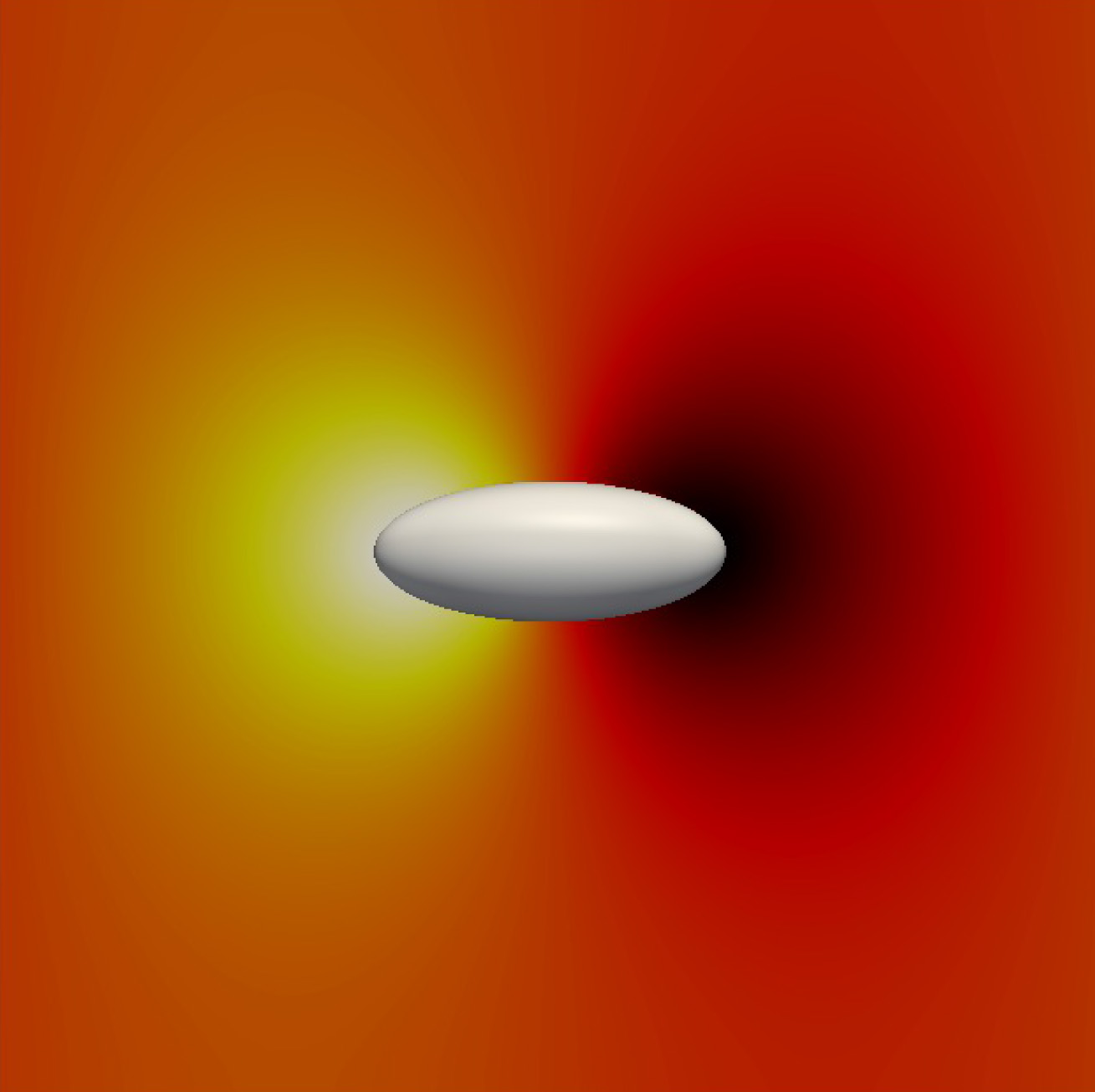} 
	\caption{(Colour Online) Particle (white) reorientation, due to the magnetic field, deforms the interface in a dipolar fashion. The colours represent the relative height of the interface: the interface is depressed on one tip of the particle (black) and raised on the other (yellow). The interface is flat in the orange/red regions. These dipolar interface deformations, called capillary charges, lead to capillary interactions between many particles at an interface, providing a route to assemble particles at fluid-fluid interfaces.}
	\label{img:dipolar_deformations}
\end{figure}

Fig~\ref{plot:fitted} shows that when the interface area removed by the particle that includes interface deformations, $A_{rm,d}$, is taken into account, the tilt-angles observed in our LB simulations are reproduced well. We note that we were unable to obtain a fit for $\alpha=1.5$, and that the critical dipole-field strength, $B_c$, is extremely sensitive to the fitting parameters: an adjustment of the fitting parameters by just $1\%$, which is smaller than errors in the measurements of $A_{rm,d}$, produces a $10\%$ difference in the critical dipole-field strength. Hence, our model in Eq.~\eqref{newfreeen}, when combined with our numerical estimate of the area, is not accurate enough to predict the exact critical dipole-field strength. We estimate the error in the theoretical predictions of the dipole-field strength, $B_c$, shown in Fig~\ref{plot:fitted} to be at least $20\%$.
However, for dipole-field strengths less than the critical dipole-field strength, $B < B_c$, the tilt-angle is robust to larger changes in the fitting parameters and the model in Eq.~\eqref{newfreeen} describes our simulation data well. \\\indent
Considering the evidence presented in Fig.~\ref{plot:interface_deform_plot} and Fig.~\ref{plot:fitted}, we conclude that the observed differences between theoretical predictions and our simulation data in Fig.~\ref{plot:theory_vs_data} can be accounted for purely by considering the interface deformations that arise due to the particle's reorientation in response to the external field.\\\indent
We have shown above that the tilt angle induced by the magnetic field features a distinctive variation with the particle aspect ratio, $\alpha$. We find that, before the first-order transition, all these data collapse into a single curve when the dipole-field strength and the tilt angle are normalized and multiplied by $\alpha$, respectively (Fig.~\ref{plot:alpha_scaling_law}). Interestingly, we did not find an $\alpha$ scaling law in the theoretical model presented in Eq.~\eqref{eq:model}, and so a convincing physical explanation of this observation is still lacking.\\\indent
The interface deformations that arise due to the particle reorienting with respect to the external field are dipolar in nature, as shown from the side in Fig.~\ref{imgs:interface_deforms} and from above in Fig.~\ref{img:dipolar_deformations}. These interface deformations are called capillary charges.~\cite{kralchevsky_capillary_1994} If there is more than one particle adsorbed at an interface under the influence of an external magnetic field, the capillary charges will interact, leading to attractive and repulsive forces as well as torques, causing the particles to aggregate and order, giving rise to novel interface structures.

\section{Conclusions}
\label{sec:conclusions}

We simulated ellipsoidal particles adsorbed at a liquid-liquid interface influenced by an external magnetic field using lattice Boltzmann (LB) simulations, which include the effect of interface deformations arising from the local fulfilment of Young's equation around the particle. These deformations arise in the continuum limit, where the colloidal particle adsorbed at the liquid-liquid interface is much larger than the surrounding fluid solvent particles.\\\indent
LB simulations provide further evidence that first-order orientation transition predicted by~\citet{bresme_orientational_2007} exists and confirm the qualitative predictions of the model, namely, that the transition depends on the particle aspect ratio, $\alpha$. However, we found quantitative differences between the tilt angles predicted by the thermodynamic model and our numerical LB simulations. These differences are in contrast with the good agreement found in atomistic simulations of small nanoparticles.~\cite{bresme_computer_2008}\\\indent 
We showed that interface deformation significantly contributes to the observed tilt angle for a given dipole-field strength and accounts for the observed deviations between LB simulations and the planar interface approximation adopted by Bresme and Faraudo.~\cite{bresme_orientational_2007} Interestingly, the latter approximation is in quantitative agreement with results obtained for small particles using Monte-Carlo simulations,~\cite{bresme_orientational_2007,bresme_computer_2008} suggesting that these deformations become less relevant for nanoparticles.~\cite{bresme_orientational_2007,bresme_computer_2008} 
Our work thus uncovers a rich physical behaviour where the influence of interface deformation depends on particle size. In the large particle limit these deformations \textit{must} be considered explicitly in order to obtain quantitative predictions of tilt angles.\\\indent
We have discovered a scaling law that reproduces the orientation behaviour of ellipsoidal particles under the influence of an external magnetic field. Renormalization of the data by the particle aspect ratio enables us to represent all the results for tilt angles \textit{vs}. dipole-field strength with a single master curve. This result is of particular relevance to future theoretical and experimental work, as it is possible to map the orientation response of particles of different aspect-ratios into a single measurement of a particular system.\\\indent
Finally, we showed that the particle tilting deforms the
interface in a dipolar manner, creating capillary charges which lead to
capillary interactions between particles. These capillary interactions should be
relevant for the self-assembly of many-particles adsorbed at fluid-fluid
interfaces, which could find applications in e.g.~colloid-liquid crystal
mixtures~\cite{meeker_colloidliquid-crystal_2000} and
photonics.~\cite{kim_self-assembled_2011} We will address this effect in a forthcoming work.  We hope that our paper will motivate new experiments to investigate the physical behaviour of anisotropic magnetic particles at fluid-fluid interfaces.

\begin{acknowledgments}
GBD \& PVC are grateful to EPRSC and Fujitsu Laboratories Europe for funding GBD's PhD studentship. JH acknowledges financial support from NWO/STW (VIDI grant 10787 of J. Harting). FB acknowledges EPSRC for the award of a Leadership Fellowship. TK thanks the University of Edinburgh for the award of a Chancellor's Fellowship.
Our work originally made use of HECToR and latterly of ARCHER, the United Kingdom’s national high-performance computing service, via CPU allocations provided through EPSRC grants ``Large Scale Lattice Boltzmann for Biocolloidal Systems” (EP/I034602/1), ``2020 Science" (http://www.2020science.net/; EP/I017909/1), and ``UK Consortium on Mesoscale Engineering Sciences"  (EP/L00030X/1). This research was also partially supported by the EU-FP7 CRESTA grant (http://www.cresta-project.eu/; Grant No. 287703). We acknowledge interesting discussions with B. Newton and M. Buzza.
\end{acknowledgments}


\end{document}